\documentclass[smallextended,final]{svjour3}
\journalname{General Relativity and Gravitation}

\usepackage{amsmath}
\usepackage{amssymb}
\usepackage[usenames,dvipsnames]{xcolor}
\usepackage[normalem]{ulem}
\usepackage[labelfont=bf,font=small]{caption}
\usepackage{graphicx}
\newcommand{\half}{{\tfrac{1}{2}}}

\begin{document} 


\title{Scalar, Vector and Tensor Harmonics on the Three-Sphere}
\titlerunning{Harmonics on the Three-Sphere}
\authorrunning{Lee Lindblom, et al.}

\author{Lee Lindblom \and Nicholas W. Taylor \and Fan Zhang}
\institute{L. Lindblom
          \at  {Center for Astro. \& Space Sci., Univ.
               of California at San Diego, La Jolla, CA 92093 U.S.A.
          \and Center for Comput. Math., Univ.
               of California at San Diego, La Jolla, CA 92093 U.S.A.
               \\\email{llindblom@ucsd.edu}}
          \and N. W. Taylor
          \at  {Department of Physics, Cornell University, Ithaca, NY 14853
                U.S.A.
                \\\email{nwt2@cornell.edu}}
          \and F. Zhang
          \at  {Grav. Wave \& Cosmology Lab., Dept. of Astron.,
               Beijing Normal Univ., Beijing 100875 China
          \and Dept. of Phys. and Astron., West Virginia Univ.,
               Morgantown, WV 26506 U.S.A.
          \\\email{fnzhang@bnu.edu.cn}}}
\date{}

\maketitle

\vspace{2.cm}

\begin{abstract}
Scalar, vector and tensor harmonics on the three-sphere were
introduced originally to facilitate the study of various problems in
gravitational physics. These harmonics are defined as eigenfunctions of 
the covariant Laplace operator which satisfy certain divergence and 
trace identities, and ortho-normality conditions. This paper provides a
summary of these properties, along with a new notation that simplifies
and clarifies some of the key expressions.  Practical methods are
described for accurately and efficiently computing these harmonics
numerically, and test results are given that illustrate how well the
analytical identities are satisfied by the harmonics computed
numerically in this way.
\end{abstract}



\section{Introduction}
\label{s:Introduction}

Scalar, vector and tensor harmonics have been introduced as part of
investigations of a variety of gravitational physics problems in
spaces with the topology of the three-sphere, $\mathbb{S}^3$.  These
include studies of the dynamics of nearly homogeneous and isotropic
cosmological models~\cite{Lifshitz1963,Halliwell1985,Lindblom2013},
the properties of quantum field theories in these
spaces~\cite{Adler1973,Adler1977}, and the dynamics of homogeneous
collapsing stellar models~\cite{Gerlach1978}.  In addition to these
application-oriented studies, a number of more mathematically focused
investigations of the properties of these harmonics have been reported
in the literature.  These include the analysis of these harmonics as
eigenfunctions of the covariant Laplace operator on
$\mathbb{S}^3$~\cite{Sandberg1978,Rubin1984,Rubin1985,LachiezeRey2005,BenAchour2016},
and as representations of the Lie groups $SU(2)$~\cite{Jantzen1978}
and $SO(4,1)$~\cite{Higuchi1991}.

Our intent here is to provide a concise summary of some of the most
useful properties of these harmonics, along with practical methods for
evaluating them numerically. Various incompatible notations for these
harmonics have been introduced and used in the references cited above.
We attempt to simplify and clarify this situation by introducing a new
uniform notation for the scalar, vector and tensor harmonics that is
analogous to notation commonly used for the more familiar harmonics on
the two-sphere, $\mathbb{S}^2$.  We summarize the useful analytical
identities satisfied by these harmonics, including their covariant
Laplace operator eigenvalue equations, their trace and divergence
identities, and their integral ortho-normality properties.  Although
we have independently verified all of these identities analytically,
we do not attempt to derive or present the proofs of them here.  These
properties have all been derived previously in the works cited above.
We do, however, present straightforward methods for evaluating these
harmonics numerically, and we present numerical tests that demonstrate
accuracy and convergence for our implementation of these numerical
methods.  These tests illustrate how well the various harmonic
identities are satisfied as functions of the numerical resolution used
to evaluate them, and also as functions of the order and the tensor
rank of the harmonics.  Section~\ref{s:ScalarHarmonics} summarizes the
properties of the scalar harmonics, Sec.~\ref{s:VectorHarmonics}
summarizes the vector and anti-symmetric second-rank tensor harmonics,
Sec.~\ref{s:TensorHarmonics} summarizes the symmetric second-rank
tensor harmonics, and finally Sec.~\ref{s:NumericalTests} describes
the results of our numerical tests that evaluate and illustrate the
accuracy of our numerical methods.

\section{Scalar Harmonics}
\label{s:ScalarHarmonics}

We use the notation $Y^{k\ell m}$ to denote the scalar harmonics on
$\mathbb{S}^3$.  The integers $k$, $\ell$, and $m$ with $k\ge \ell\geq 0$ and
$\ell\geq m \geq -\ell$ indicate the order of the harmonic. These
harmonics are eigenfunctions of the covariant Laplace operator:
\begin{eqnarray}
  \nabla^a\nabla_a Y^{k\ell m} = -\frac{k(k+2)}{R_3^2} Y^{k\ell m},
  \label{e:ScalarEigenvalueEq}
\end{eqnarray} 
where $R_3$ is the radius, and $\nabla_a$ is the covariant derivative
associated with the round metric on $\mathbb{S}^3$.  It is often useful to
express this metric in terms of spherical coordinates
$x^a=(\chi,\theta,\varphi)$:
\begin{eqnarray}
ds^2&=&g_{ab}dx^adx^b,\nonumber\\ 
&=& R_3^2\left[d\chi^2+\sin^2\chi\left(
d\theta^2+\sin^2\theta\, d\varphi^2\right)\right].
\end{eqnarray}
The $\mathbb{S}^3$ scalar harmonics can be expressed in terms of the standard
$\mathbb{S}^2$ scalar harmonics $Y^{\ell m}$ by defining the functions
$H^{k\ell}(\chi)$: 
\begin{eqnarray}
Y^{k\ell m}(\chi,\theta,\varphi)=H^{k\ell}(\chi)Y^{\ell m}(\theta,\varphi).
\label{e:HklDef}
\end{eqnarray}
The $H^{k\ell}(\chi)$ are determined by inserting Eq.~(\ref{e:HklDef}) into
Eq.~(\ref{e:ScalarEigenvalueEq}), to obtain the following ordinary
differential equation:
\begin{eqnarray}
  0&=& \frac{d^2H^{k\ell}}{d\chi^2}+2\cot\chi\frac{dH^{k\ell}}{d\chi}
  +\left[k(k+2)-\ell(\ell+1)\csc^2\chi\right] H^{k\ell}.\qquad
\end{eqnarray}
This differential equation can be solved most conveniently
by introducing $C^{k\ell}(\chi)$ such that 
\begin{eqnarray}
H^{k\ell}(\chi)=\sin^\ell\!\chi\,C^{k\ell}(\chi).
\label{e:CklDef}
\end{eqnarray}
These $C^{k\ell}$ (which are proportional to the Gegenbauer
polynomials) are given, for $k=\ell$ and $k=\ell+1$, by the
expressions
\begin{eqnarray} 
C^{\ell\ell}&=& (-1)^{\ell+1}2^{\ell}\ell\kern 0.1em!\,
\sqrt{\frac{2(\ell+1)}{\pi(2\ell+1)!}},\qquad\\
C^{\ell+1\,\ell} &=& \sqrt{2(\ell+2)}\cos\chi\, C^{\ell\ell}.
\end{eqnarray}
The solutions for $k>\ell+1$ are determined by
applying the recursion relation
\begin{eqnarray}
&&\!\!\!\!\!\!\!\!
C^{k+2\,\ell} = 
2\cos\chi \sqrt{\frac{(k+3)(k+2)}{(k+3+\ell)(k+2-\ell)}}\,C^{k+1\,\ell} 
\qquad\nonumber\\
&&\qquad\qquad\quad
-\sqrt{\frac{(k+3)(k+2+\ell)(k+1-\ell)}
{(k+1)(k+3+\ell)(k+2-\ell)}}\,C^{k\ell},\qquad\label{e:CRecursion}
\end{eqnarray}
iteratively starting with the $k=\ell$ and $k=\ell+1$ solutions given
above.  This allows the $Y^{k\ell m}$ to be determined
numerically in a straightforward way as functions of the spherical
coordinates $(\chi,\theta,\varphi)$ using Eqs.~(\ref{e:HklDef}) and
(\ref{e:CklDef})--(\ref{e:CRecursion}).

The $\mathbb{S}^3$ scalar harmonics $Y^{k\ell m}$, defined and normalized as
above, satisfy the following ortho-normality conditions:
\begin{eqnarray}
\delta^{\,kk'}\delta^{\,\ell\ell'}\delta^{\,mm'}
=\frac{1}{R_3^3}\int Y^{k\ell m}\, Y^{*k'\ell'm'}\sqrt{g}\,d^{\,3}x.
\label{e:ScalarOrthoNormal}
\end{eqnarray}
Any scalar function $S$ on the three-sphere can be written as an
expansion in terms of these scalar harmonics:
\begin{eqnarray}
S=\sum_{k=0}^\infty\sum_{\ell=0}^{k}\sum_{m=-\ell}^\ell S^{k\ell m}\,Y^{k\ell m}.
\end{eqnarray}
The ortho-normality relations in Eq.~(\ref{e:ScalarOrthoNormal}) imply the
following integral expressions for the expansion coefficients
$S^{k\ell m}$,
\begin{eqnarray}
S^{k\ell m}=\frac{1}{R_3^3}\int S \,Y^{*k\ell m} \sqrt{g}\,d^{\,3}x.
\end{eqnarray}

We find it useful in our numerical work on the three-sphere to
represent fields in coordinates other than the standard spherical
coordinate system.  Let the transformation law $x^b=x^b(\bar x^a)$,
denote the transformation between spherical coordinates
$x^a=(\chi,\theta,\varphi)$ and some other useful coordinates $\bar
x^a$.  Given these coordinate transformation laws, it is
straightforward to determine $Y^{k\ell m}(\bar x^a)$ simply by
composing the standard spherical coordinate representation described
above, $Y^{k\ell m}(x^b)$, with the coordinate transformation law
$x^b(\bar x^a)$ to obtain $Y^{k\ell m}(\bar x^b)=Y^{k\ell m}[x^a(\bar
  x^b)]$.

\section{Vector Harmonics}
\label{s:VectorHarmonics}

The vector harmonics on $\mathbb{S}^3$ are completely determined by the scalar
harmonics $Y^{k\ell m}$ and their derivatives.  In particular the
three classes of vector harmonics, which we denote $Y^{k\ell
  m}_{(A\,)a}$ for ${\scriptstyle A}=0,1,2$,
are defined by
\begin{eqnarray}
Y^{k\ell m}_{(0)\,a}&=& \frac{R_3}{\sqrt{k(k+2)}}\nabla_aY^{k\ell m},
\label{e:Yklm(0)i}\\
Y^{k\ell m}_{(1)\,a}&=& \frac{R_3^2}{\sqrt{\ell(\ell+1)}}
\epsilon_{a}{}^{bc}\nabla_bY^{k\ell m}\nabla_c\cos\chi,
\label{e:Yklm(1)i}\\
Y^{k\ell m}_{(2)\,a}&=& \frac{R_3}{k+1}\epsilon_a{}^{bc}\nabla_bY^{k\ell m}_{(1)\,c},
\label{e:Yklm(2)i}
\end{eqnarray}
where $\epsilon_{abc}$ is the totally antisymmetric tensor volume
element, which satisfies $\nabla_d\epsilon_{abc}=0$.  All the vector
harmonics vanish identically for $k=0$, and the harmonics $Y^{k\ell
  m}_{(1)\,a}$ and $Y^{k\ell m}_{(2)\,a}$ are not well defined for
$\ell=0$.  Thus the vector harmonics are defined only for $k\geq
k^{V(A)}_{\mathrm{min}}$ and $\ell\geq \ell^{V(A)}_{\mathrm{min}}$,
where the values of $k^{V(A)}_{\mathrm{min}}$ and
$\ell^{V(A)}_{\mathrm{min}}$ are given in Table~\ref{t:TableI}.
\begin{table}[h]
  \begin{center}
\begin{tabular}{c c c c c}
$\scriptstyle A$ & $k_\mathrm{min}^{V(A)}$ & $\ell_\mathrm{min}^{V(A)}$ &
$k_\mathrm{min}^{T(A)}$ & $\ell_\mathrm{min}^{T(A)}$ \\
\hline
0 & 1 & 0 & 0 & 0 \\ 
1 & 1 & 1 & 2 & 1 \\
2 & 1 & 1 & 2 & 1 \\
3 & - & - & 2 & 0 \\
4 & - & - & 2 & 2 \\
5 & - & - & 2 & 2 \\
\end{tabular}
\end{center}
\caption{Minimum values of the harmonic order parameters $k\geq
  k_\mathrm{min}$ and $\ell\geq \ell_\mathrm{min}$ for the various
  classes of vector and tensor harmonics.}
\label{t:TableI}
\end{table}

The vector harmonics $Y^{k\ell m}_{(A)\,a}$ are eigenfunctions of the
covariant Laplace operator on $\mathbb{S}^3$, which satisfy the following
eigenvalue equations,
\begin{eqnarray}
  \nabla^b\nabla_bY^{k\ell m}_{(0)\,a}&=& \frac{2-k(k+2)}{R_3^2}Y^{k\ell m}_{(0)\,a},
  \label{e:EigVecY0}\\
  \nabla^b\nabla_bY^{k\ell m}_{(1)\,a}&=& \frac{1-k(k+2)}{R_3^2}Y^{k\ell m}_{(1)\,a},
  \label{e:EigVecY1}\\
  \nabla^b\nabla_bY^{k\ell m}_{(2)\,a}&=& \frac{1-k(k+2)}{R_3^2}Y^{k\ell m}_{(2)\,a}.
  \label{e:EigVecY2}
\end{eqnarray}
These vector harmonics also satisfy the following divergence identities,
\begin{eqnarray}
  \nabla^aY^{k\ell m}_{(0)\,a}&=& -\frac{\sqrt{k(k+2)}}{R_3}Y^{k\ell m},
  \label{e:DivVecY0}\\
\nabla^aY^{k\ell m}_{(1)\,a}&=& 0\label{e:DivVecY1},\\
\nabla^aY^{k\ell m}_{(2)\,a}&=& 0\label{e:DivVecY2}.
\end{eqnarray}

The vector harmonics defined in Eqs.~(\ref{e:Yklm(0)i})--(\ref{e:Yklm(2)i})
have been normalized so they satisfy the following ortho-normality
relations,
\begin{eqnarray}
\delta_{(A)(B)}\delta^{\,kk'}\delta^{\,\ell\ell'}\delta^{\,mm'}=
\frac{1}{R_3^3}\int\!\! 
g^{ab}\,Y^{\,k\ell m}_{(A)\,a}\,Y^{\,*k'\ell' m'}_{(B)\,b}\!\sqrt{g}\,
d^{\,3}x,\!\!\!\!\!\nonumber\\
\label{e:VectorOrthoNormal}
\end{eqnarray}
for $k,k'\geq k^{V(A)}_\mathrm{min}$ and $\ell,\ell'\geq
\ell^{V(A)}_\mathrm{min}$.  It is often useful and convenient to
express vector fields $V^a$ as expansions in terms of these vector
harmonics:
\begin{eqnarray}
  V^a=g^{ab}\sum_{A=0}^{2}\sum_{k=k^{V(A)}_\mathrm{min}}^\infty
  \sum_{\ell=\ell^{V(A)}_\mathrm{min}}^{k}\sum_{m=-\ell}^\ell 
V^{k\ell m}_{(A)}
\,Y^{k\ell m}_{(A)\,b}.\qquad
\end{eqnarray}
The ortho-normality relations, Eq.~(\ref{e:VectorOrthoNormal}),
imply that the expansion coefficients $V^{k\ell m}_{(A)}$ are simply
the integral projections of the vector field onto the corresponding
vector harmonics:
\begin{eqnarray}
V^{k\ell m}_{(A)}=\frac{1}{R_3^3}\int V^a \,Y^{*k\ell m}_{(A)\,a} \sqrt{g}\,d^{\,3}x.
\label{e:VectorExpansionCoefficients}
\end{eqnarray}
We point out that the new notation we have introduced here for these
harmonics makes it much simpler to express the ideas contained in
Eqs.~(\ref{e:VectorOrthoNormal})--(\ref{e:VectorExpansionCoefficients})
than it would have been using previous notations.

The expressions for the vector harmonics given in
Eqs.~(\ref{e:Yklm(0)i})--(\ref{e:Yklm(2)i}) are covariant, so it is
straightforward to evaluate them in any convenient coordinates.  To
evaluate them numerically, we start by computing the scalar harmonics
$Y^{k\ell m}$ numerically on a grid of points in the chosen
coordinates using the methods described in
Sec.~\ref{s:ScalarHarmonics}.  The gradients of the scalar harmonics
are then evaluated on this grid using any convenient numerical method
(e.g, finite difference or pseudo-spectral).  Finally these gradients
are combined algebraically using the expressions in
Eqs.~(\ref{e:Yklm(0)i})--(\ref{e:Yklm(2)i}) to determine the vector
harmonics on that grid of points.

Anti-symmetric tensor fields, $W_{ab}=-W_{ba}$, in three-dimensions
are dual to vector fields: for every $W_{ab}$ there exists a vector
field $V^a$ such that $W_{ab}=\epsilon_{abc}V^c$.  Thus arbitrary
anti-symmetric tensor fields on $\mathbb{S}^3$ can be expressed in terms of the
vector harmonics:
\begin{eqnarray}
W_{ab}=\epsilon_{ab}{}^c 
\sum_{A=0}^{2}\sum_{k=k^{V(A)}_\mathrm{min}}^\infty
  \sum_{\ell=\ell^{V(A)}_\mathrm{min}}^{k}\sum_{m=-\ell}^\ell V^{k\ell m}_{(A)}
\,Y^{k\ell m}_{(A)\,c}.\qquad
\end{eqnarray}
The expansion coefficients $V^{k\ell m}_{(A)}$ are given as before by
Eq.~(\ref{e:VectorExpansionCoefficients}).

\section{Tensor Harmonics}
\label{s:TensorHarmonics}

The symmetric second-rank tensor harmonics on $\mathbb{S}^3$ are determined by
the scalar and vector harmonics defined in
Secs.~\ref{s:ScalarHarmonics} and \ref{s:VectorHarmonics}, and their
derivatives.  There are six classes of these harmonics, which we
denote $Y^{k\ell m}_{(A)\,ab}$ for ${\scriptstyle A} = 0,1,2,3,4,5$,
defined as,
\begin{eqnarray}
&&
Y^{k\ell m}_{(0)\,ab}= \frac{1}{\sqrt{3}}Y^{k\ell m}g_{ab},
\label{e:Yklm(0)ij}\\
&&
Y^{k\ell m}_{(1)\,ab}= \frac{R_3}{\sqrt{2(k-1)(k+3)}}
\left(\nabla_aY^{k\ell m}_{(1)\,b}
+\nabla_bY^{k\ell m}_{(1)\,a}\right),
\label{e:Yklm(1)ij}\\
&&
Y^{k\ell m}_{(2)\,ab}= \frac{R_3}{\sqrt{2(k-1)(k+3)}}
\left(\nabla_aY^{k\ell m}_{(2)\,b}
+\nabla_bY^{k\ell m}_{(2)\,a}\right),
\label{e:Yklm(2)ij}\\
&&
Y^{k\ell m}_{(3)\,ab}= \frac{\sqrt{3}R_3}{\sqrt{2(k-1)(k+3)}}
\Biggl(\nabla_a Y^{k\ell m}_{(0)\,b}
+\frac{\sqrt{k(k+2)}}{\sqrt{3}R_3}Y^{k\ell m}_{(0)\,ab}\Biggr),\\
\label{e:Yklm(3)ij}
&&
Y^{k\ell m}_{(4)ab}= R_3\sqrt{\frac{(\ell-1)(\ell+2)}{2k(k+2)}}\biggl\{\half 
E^{k\ell}\left( \nabla_a F^{\ell m}_{b}
+\nabla_b F^{\ell m}_{a}\right)\nonumber\\
&& \qquad+ 
\csc^2\!\chi\left[\tfrac{1}{2}(\ell-1)\cos\chi\, E^{k\ell}+C^{k\ell}
\right]
\left(F^{\ell m}_{a}\nabla_b\cos\chi+ F^{\ell m}_{b}\nabla_a\cos\chi\right)
\biggr\},\quad
\label{e:Yklm(4)ij}\\
&&
Y^{k\ell m}_{(5)\,ab}= \frac{R_3}{2(k+1)}
\left(\epsilon_a{}^{ce}\nabla_cY^{k\ell m}_{(4)\,eb} 
+\epsilon_b{}^{ce}\nabla_cY^{k\ell m}_{(4)\,ea}\right) .\qquad
\label{e:Yklm(5)ij}
\end{eqnarray}
The quantities $E^{k\ell}$ and $F^{\ell m}_a$ that appear
in Eq.~(\ref{e:Yklm(4)ij}) are given by
\begin{eqnarray}
E^{k\ell}&=& -\frac{2\csc^{\ell+1}\!\chi}{(\ell-1)(\ell+2)}
\frac{d}{d\chi}\left(\sin^2\chi H^{kl}\right),\\
F^{\ell m}_a&=&\frac{R_3^2}{\sqrt{\ell(\ell+1)}}
\epsilon_a{}^{bc}\,\nabla_b \left(\sin^\ell\!\chi\,
Y^{\ell m}\right)\nabla_c\cos\chi,\label{e:FlmDef}
\end{eqnarray}
where the $H^{k\ell}$ are given in Eq.~(\ref{e:HklDef}).
The functions $E^{k\ell}(\chi)$ can be computed numerically
for $k=\ell$ from the expression,
\begin{eqnarray}
E^{\ell\ell}(\chi)&=&-\frac{2\cos\chi}{\ell-1}C^{\ell\ell}(\chi),
\end{eqnarray}
and for $k>\ell$ from
\begin{eqnarray}
E^{k\ell}(\chi)&=&-\frac{2(k+2)\cos\chi}{(\ell-1)(\ell+2)}C^{k\ell}(\chi) 
+\frac{2\sqrt{(k+1)(k-\ell)(k+\ell+1)}}{(\ell-1)(\ell+2)\sqrt{k}}
C^{k-1\,\ell}(\chi),\qquad
\end{eqnarray}
where the $C^{k\,\ell}(\chi)$ are given in Eq.~(\ref{e:CklDef}).
The tensor harmonics $Y^{k\ell m}_{(A)\,ab}$ are only defined for
$k\geq k^{T(A)}_\mathrm{min}$ and $\ell\geq \ell^{T(A)}_\mathrm{min}$,
where the minimum values $k^{T(A)}_\mathrm{min}$ and
$\ell^{T(A)}_\mathrm{min}$ are listed in Table~\ref{t:TableI} for each
class of harmonics.

The symmetric second-rank tensor harmonics $Y^{k\ell m}_{(A)\,ab}$ are
eigenfunctions of the covariant Laplace operator on $\mathbb{S}^3$ that satisfy
the following eigenvalue equations,
\begin{eqnarray}
  \nabla^n\nabla_nY^{k\ell m}_{(0)\,ab}&=& -\frac{k(k+2)}{R_3^2}Y^{k\ell m}_{(0)\,ab},
  \label{e:EigTensorY0}\\
  \nabla^n\nabla_nY^{k\ell m}_{(1)\,ab}&=& \frac{5-k(k+2)}{R_3^2}Y^{k\ell m}_{(1)\,ab},
  \label{e:EigTensorY1}\\
  \nabla^n\nabla_nY^{k\ell m}_{(2)\,ab}&=& \frac{5-k(k+2)}{R_3^2}Y^{k\ell m}_{(2)\,ab},
  \label{e:EigTensorY2}\\
  \nabla^n\nabla_nY^{k\ell m}_{(3)\,ab}&=& \frac{6-k(k+2)}{R_3^2}Y^{k\ell m}_{(3)\,ab},
  \label{e:EigTensorY3}\\
  \nabla^n\nabla_nY^{k\ell m}_{(4)\,ab}&=& \frac{2-k(k+2)}{R_3^2}Y^{k\ell m}_{(4)\,ab},
  \label{e:EigTensorY4}\\
  \nabla^n\nabla_nY^{k\ell m}_{(5)\,ab}&=& \frac{2-k(k+2)}{R_3^2}Y^{k\ell m}_{(5)\,ab}.
  \label{e:EigTensorY5}
\end{eqnarray}
These harmonics also satisfy the following divergence
identities,
\begin{eqnarray}
\nabla^aY^{k\ell m}_{(0)\,ab}&=& \frac{\sqrt{k(k+2)}}{\sqrt{3}R_3}
Y^{k\ell m}_{(0)\,b},\label{e:DivTensorY0}\\
\nabla^aY^{k\ell m}_{(1)\,ab}&=& 
-\frac{\sqrt{(k-1)(k+3)}}{\sqrt{2}R_3}Y^{k\ell m}_{(1)\,b},\label{e:DivTensorY1}\\
\nabla^aY^{k\ell m}_{(2)\,ab}&=& 
-\frac{\sqrt{(k-1)(k+3)}}{\sqrt{2}R_3}Y^{k\ell m}_{(2)\,b},\label{e:DivTensorY2}\\
\nabla^aY^{k\ell m}_{(3)\,ab}&=& -\frac{\sqrt{2(k-1)(k+3)}}{\sqrt{3}R_3}
Y^{k\ell m}_{(0)\,b},\label{e:DivTensorY3}\\
\nabla^aY^{k\ell m}_{(4)\,ab}&=& 0,\label{e:DivTensorY4}\\
\nabla^aY^{k\ell m}_{(5)\,ab}&=& 0\label{e:DivTensorY5},
\end{eqnarray}
and the following trace conditions,
\begin{eqnarray}
  g^{ab}Y^{k\ell m}_{(A)\,ab} &=&\left\{
  \begin{array}{c l}
    \sqrt{3}Y^{k\ell m}, & \qquad {\scriptstyle A}=0, \\
    0, & \qquad 1\le {\scriptstyle A} \le 5. \\
  \end{array}
  \right.\label{e:TraceIdentities}
\end{eqnarray}

The symmetric second-rank tensor harmonics $Y^{k\ell m}_{(A)\,ab}$
satisfy the following ortho-normality conditions,
\begin{eqnarray}
  &&\delta_{(A)(B)}\delta^{\,kk'}\delta^{\,\ell\ell'}\delta^{\,mm'}=
\frac{1}{R_3^3}\int g^{ac}g^{bd}\,Y^{\,k\ell m}_{(A)\,ab}\,Y^{\,*k'\ell' m'}_{(B)\,cd}
\!\sqrt{g}\,
d^{\,3}x,\qquad
\label{e:TensorOrthoNormal}
\end{eqnarray}
for $k,k'\geq k^{T(A)}_\mathrm{min}$ and $\ell,\ell'\geq
\ell^{T(A)}_\mathrm{min}$.  It is often useful to express symmetric
second-rank tensor fields $T_{ab}$ on $\mathbb{S}^3$ as expansions in terms of
these tensor harmonics:
\begin{eqnarray}
  T_{ab}=\sum_{A=0}^{5}\sum_{k=k^{T(A)}_\mathrm{min}}^\infty
  \sum_{\ell=\ell^{T(A)}_\mathrm{min}}^{k}\sum_{m=-\ell}^\ell 
T^{k\ell m}_{(A)}
\,Y^{k\ell m}_{(A)\,ab}.
\end{eqnarray}
The ortho-normality relations, Eq.~(\ref{e:TensorOrthoNormal}),
make it easy to express the expansion coefficients $T^{k\ell m}_{(A)}$ 
as the projections of the tensor $T_{ab}$ onto the tensor harmonics:
\begin{eqnarray}
T^{k\ell m}_{(A)}=\frac{1}{R_3^3}\int g^{ac}g^{bd}
T_{ab} \,Y^{*k\ell m}_{(A)\,cd}\, \sqrt{g}\,d^{\,3}x.
\label{e:TensorCoefficients}
\end{eqnarray}
We point out again that the notation used here makes the fundamental
identities in
Eqs.~(\ref{e:TensorOrthoNormal})--(\ref{e:TensorCoefficients}) much
simpler to express than they would be with earlier notations.

The expressions for the tensor harmonics given in
Eqs.~(\ref{e:Yklm(0)ij})--(\ref{e:Yklm(5)ij}) are covariant, so it is
straightforward to evaluate them in any convenient coordinates.  To
evaluate them numerically, we begin by evaluating the scalar and
vector harmonics numerically on a grid of points in the chosen
coordinates using the methods described in
Secs.~\ref{s:ScalarHarmonics} and \ref{s:VectorHarmonics}.  Next we
evaluate the co-vectors $F^{\ell m}_a$ defined in Eq.~(\ref{e:FlmDef})
and $\nabla_a\cos\chi$ numerically on this same grid.  Finally we
compute the covariant gradients of the vector harmonics numerically on
this grid, and combine the various terms algebraically to determine
the tensor harmonics using the expressions in
Eqs.~(\ref{e:Yklm(0)ij})--(\ref{e:Yklm(5)ij}).

Up to normalizations, our expressions for the tensor harmonics
Eqs.~(\ref{e:Yklm(0)ij})--(\ref{e:Yklm(5)ij}) are equivalent to those
given in Ref.~\cite{Sandberg1978} (using very different notation).
Our expression for Eq.~(\ref{e:Yklm(4)ij}) has been re-written however
in a form that makes it easier to evaluate numerically.  The analogous
expression in Ref.~\cite{Sandberg1978} includes terms that become
singular at the poles $\chi=0$ and $\chi=\pi$.  The singular behavior
in those terms cancels analytically, but that behavior makes it
difficult to evaluate them numerically with good precision.  Our
re-written expression for Eq.~(\ref{e:Yklm(4)ij}) eliminates those
singular terms, making it much more suitable for numerical work.

\section{Numerical Tests}
\label{s:NumericalTests}

This section describes the tests we have performed to measure the
accuracy of the scalar, vector and tensor harmonics on $\mathbb{S}^3$
computed numerically using the methods outlined in
Secs.~\ref{s:ScalarHarmonics}--\ref{s:TensorHarmonics}.  To perform
these tests we have implemented these numerical methods in the SpEC
code~\cite{Pfeiffer2003,LindblomSzilagyi2011a}.  This code uses
pseudo-spectral methods for constructing the numerical grids and for
evaluating numerical derivatives and integrals of fields.
Pseudo-spectral methods converge exponentially in the number of grid
points used to represent the fields and are very efficient at
producing high accuracy results with minimal computational cost.  We
note, however, that the methods described in
Secs.~\ref{s:ScalarHarmonics}--\ref{s:TensorHarmonics} are quite
general and could be implemented using any standard numerical method
(e.g., finite difference or finite element).

The three-sphere, $\mathbb{S}^3$, is not homeomorphic to $\mathbb{R}^3$,
so it can not
be covered smoothly by a single coordinate patch.  For the tests
described here, we use a multi-cube representation of $\mathbb{S}^3$ having
eight cubic non-overlapping coordinate
patches~\cite{LindblomSzilagyi2011a} that is analogous to the
cubed-sphere representations of $\mathbb{S}^2$~\cite{Ronchi1996}.  We represent
the fields needed to compute the $\mathbb{S}^3$ harmonics on this manifold using
pseudo-spectral coordinate grids having $N$ grid points in each
direction in each of the eight coordinate patches.  The total number
of grid points used to represent each field in our tests on $\mathbb{S}^3$ is
therefore $8N^3$.  The coordinate transformation relating the standard
spherical coordinates to the multi-cube coordinates used in our tests
is given explicitly in Ref.~\cite{LindblomSzilagyi2011a}.  This
transformation allows us to evaluate the spherical coordinates
$x^a=(\chi,\theta,\varphi)$ as functions on the numerical grid. Any
function of the spherical coordinates can then be evaluated easily on
this grid using these spherical coordinate functions.  In this way the
numerical methods described in
Secs.~\ref{s:ScalarHarmonics}--\ref{s:TensorHarmonics} are used in our
tests to evaluate the scalar, vector and tensor harmonics on these
multi-cube grids.

We have developed a series of tests designed to determine how well our
numerical implementations of these harmonics on $\mathbb{S}^3$ actually work.
In particular we measure the numerical residuals obtained when
evaluating the various identities satisfied analytically by these
harmonics. The first set of residuals measures how well the eigenvalue
equations are satisfied.  Let $\mathcal{E}^{k\ell m}$ denote the
residual for the scalar harmonic eigenvalue equation given in
Eq.~(\ref{e:ScalarEigenvalueEq}):
\begin{eqnarray}
  \mathcal{E}^{k\ell m} &=& \nabla^a\nabla_a Y^{k\ell m}
  +\frac{k(k+2)}{R_3^2} Y^{k\ell m}.
\end{eqnarray}
This residual (and all the other residuals we define) should vanish
identically, so measuring its deviation from zero allows us to
evaluate the accuracy of our numerical methods quantitatively.
Analogous expressions are defined for the residuals
$\mathcal{E}^{k\ell m}_{(A)\,a}$ of the vector harmonic eigenvalue
equations from Eqs.~(\ref{e:EigVecY0})--(\ref{e:EigVecY2}), and for
the residuals $\mathcal{E}^{k\ell m}_{(A)\,ab}$ of the tensor harmonic
eigenvalue equations from
Eqs.~(\ref{e:EigTensorY0})--(\ref{e:EigTensorY5}).  We measure how
well these identities are satisfied by evaluating the norm
$\|Q\|_2$ of these quantities, defined as
\begin{eqnarray}
\left(  \|\mathcal{E}^{k\ell m}\|_2\right)^2&=& \frac{\int \mathcal{E}^{k\ell m}
    \mathcal{E}^{k\ell m*}\sqrt{g}\,d^{\,3}x}
  {\int \sqrt{g}\,d^{\,3}x},\\
  \left(  \|\mathcal{E}^{k\ell m}_{(A)a}\|_2\right)^2&=&
  \frac{\int g^{ab}\mathcal{E}^{k\ell m}_{(A)a}
    \mathcal{E}^{k\ell m*}_{(A)b}\sqrt{g}\,d^{\,3}x}
  {\int \sqrt{g}\,d^{\,3}x},\\
  \left(  \|\mathcal{E}^{k\ell m}_{(A)ab}\|_2\right)^2&=&
  \frac{\int g^{ab}g^{cd}\mathcal{E}^{k\ell m}_{(A)ac}
    \mathcal{E}^{k\ell m*}_{(A)bd}\sqrt{g}\,d^{\,3}x}
  {\int \sqrt{g}\,d^{\,3}x},
\end{eqnarray}
for scalar, vector and tensor quantities respectively.

The second set of identities of interest to us are those for the
divergences of the vector and tensor harmonics in
Eqs.~(\ref{e:DivVecY0})--(\ref{e:DivVecY2}) and
Eqs.~(\ref{e:DivTensorY0})--(\ref{e:DivTensorY5}) respectively.  We
define the vector harmonic divergence residuals
$\mathcal{D}^{klm}_{(A)}$ as the left sides minus the right sides of
Eqs.~(\ref{e:DivVecY0})--(\ref{e:DivVecY2}).  For example
$\mathcal{D}^{klm}_{(0)}$ is given by
\begin{eqnarray}
  \mathcal{D}^{klm}_{(0)} &=&   \nabla^aY^{k\ell m}_{(0)\,a}
  +\frac{\sqrt{k(k+2)}}{R_3}Y^{k\ell m}.
\end{eqnarray}
The tensor harmonic divergence residuals $\mathcal{D}^{k\ell
  m}_{(A)\,a}$ are defined analogously from
Eqs.~(\ref{e:DivTensorY0})--(\ref{e:DivTensorY5}).  We monitor how
well these identities are satisfied by evaluating their $\|Q\|_2$
norms, as defined above.  The third set of identities of interest to
us are the trace identities for the tensor harmonics given in
Eq.~(\ref{e:TraceIdentities}).  For example $\mathcal{T}^{klm}_{(0)}$
is given by
\begin{eqnarray}
\mathcal{T}^{k\ell m}_{(0)}=g^{ab}Y^{k\ell m}_{(0)\,ab}-\sqrt{3}Y^{k\ell m},
\end{eqnarray}
with analogous expressions for the remaining $\mathcal{T}^{k\ell
  m}_{(A)}$.  As in the previous identities, we monitor how well these
are satisfied by evaluating their $\|Q\|_2$ norms.

We note that the eigenvalue residuals $\mathcal{E}^{k\ell m}$,
etc. satisfy the symmetry conditions, $\mathcal{E}^{k\ell\,+
  m}=\bigl(\mathcal{E}^{k\ell\,- m}\bigr)^*$; the divergence and
trace residuals satisfy similar conditions.  Since the norms $\|Q\|_2$
of these residuals are the same for the $-m$ harmonics as they are for
the corresponding $+m$ harmonics, it is only necessary to evaluate
them for harmonics with $m\geq 0$.

Finally we define a set of residuals that measure how well the various
harmonics satisfy the ortho-normality conditions given in
Eqs.~(\ref{e:ScalarOrthoNormal}), (\ref{e:VectorOrthoNormal}), and
(\ref{e:TensorOrthoNormal}).  For example, we define the scalar
harmonic ortho-normality residuals $S\mathcal{O}^{k\ell
  m}_{k'\ell'm'}$ from Eq.~(\ref{e:ScalarOrthoNormal}) as
\begin{eqnarray}
  S\mathcal{O}^{k\ell m}_{k'\ell'm'}&=&
  \delta^{\,kk'}\delta^{\,\ell\ell'}\delta^{\,mm'}
  -\frac{1}{R_3^3}\int Y^{k\ell m}\, Y^{*k'\ell'm'}\sqrt{g}\,d^{\,3}x.
\end{eqnarray}
We also define analogous vector harmonic ortho-normality residuals
$V\mathcal{O}^{(A)k\ell m}_{(B)k'\ell'm'}$ from
Eq.~(\ref{e:VectorOrthoNormal}), and tensor harmonics ortho-normality
residuals $T\mathcal{O}^{(A)k\ell m}_{(B)k'\ell'm'}$ from
Eq.~(\ref{e:TensorOrthoNormal}).

In addition to the residuals defined above that measure how well each
individual identity is satisfied, it is useful to define composite
residuals that measure how well classes of identities are satisfied.
Thus we define the composite scalar harmonic eigenvalue residual
$S\mathcal{E}_2$, which measures the average value of
$\|\mathcal{E}^{k\ell m}\|_2$:
\begin{eqnarray}
  (\mathcal{SE}_2)^2 =
  \frac{1}{N_{k\ell m}^\geq}\sum_{k=0}^{k_\mathrm{max}}\sum_{\ell=0}^k\sum_{
    m=0}^\ell \left(\|\mathcal{E}^{k\ell m}\|_2\right)^2,
  \label{e:SE2}
\end{eqnarray}
where $N_{k\ell
  m}^\geq=(k_\mathrm{max}+1)(k_\mathrm{max}+2)(k_\mathrm{max}+3)/6$ is
the total number of $k\ell m$ triplets with $m\geq 0$ included in the
sums.  We also define analogous composite eigenvalue residuals for the
vector and tensor cases:
\begin{eqnarray}
  (V\mathcal{E}_2)^2 &=&
\sum_{A=0}^2\sum_{k=k^{V(A)}_\mathrm{min}}^{k_\mathrm{max}}
  \sum_{\ell=\ell^{V(A)}_\mathrm{min}}^k\sum_{
    m=0}^\ell
  \frac{\bigl(\|\mathcal{E}^{k\ell m}_{(A)\,a}\|_2\bigr)^2}
       {3N_{k\ell m}^\geq-2k_\mathrm{max}-3}
  ,\label{e:VE2} \\[1em]
  (T\mathcal{E}_2)^2 &=&
\sum_{A=0}^5\sum_{k=k^{T(A)}_\mathrm{min}}^{k_\mathrm{max}}
  \sum_{\ell=\ell^{T(A)}_\mathrm{min}}^k\sum_{m=0}^\ell
  \frac{\bigl(\|\mathcal{E}^{k\ell m}_{(A)\,ab}\|_2\bigr)^2}
       {6N_{k\ell m}^\geq-8k_\mathrm{max}-12}.
       \label{e:TE2}
      \\ \nonumber
 \end{eqnarray}
The terms $2k_\mathrm{max}+3$ and $8k_\mathrm{max}+12$ that appear in
Eqs.~(\ref{e:VE2}) and (\ref{e:TE2}) respectively represent the number
of terms excluded in these sums by the lower bounds $k\geq
k^{T(A)}_\mathrm{min}$ and $\ell\geq \ell^{T(A)}_\mathrm{min}$.

\begin{figure}
  \centering
  \includegraphics[width=0.31\textwidth]{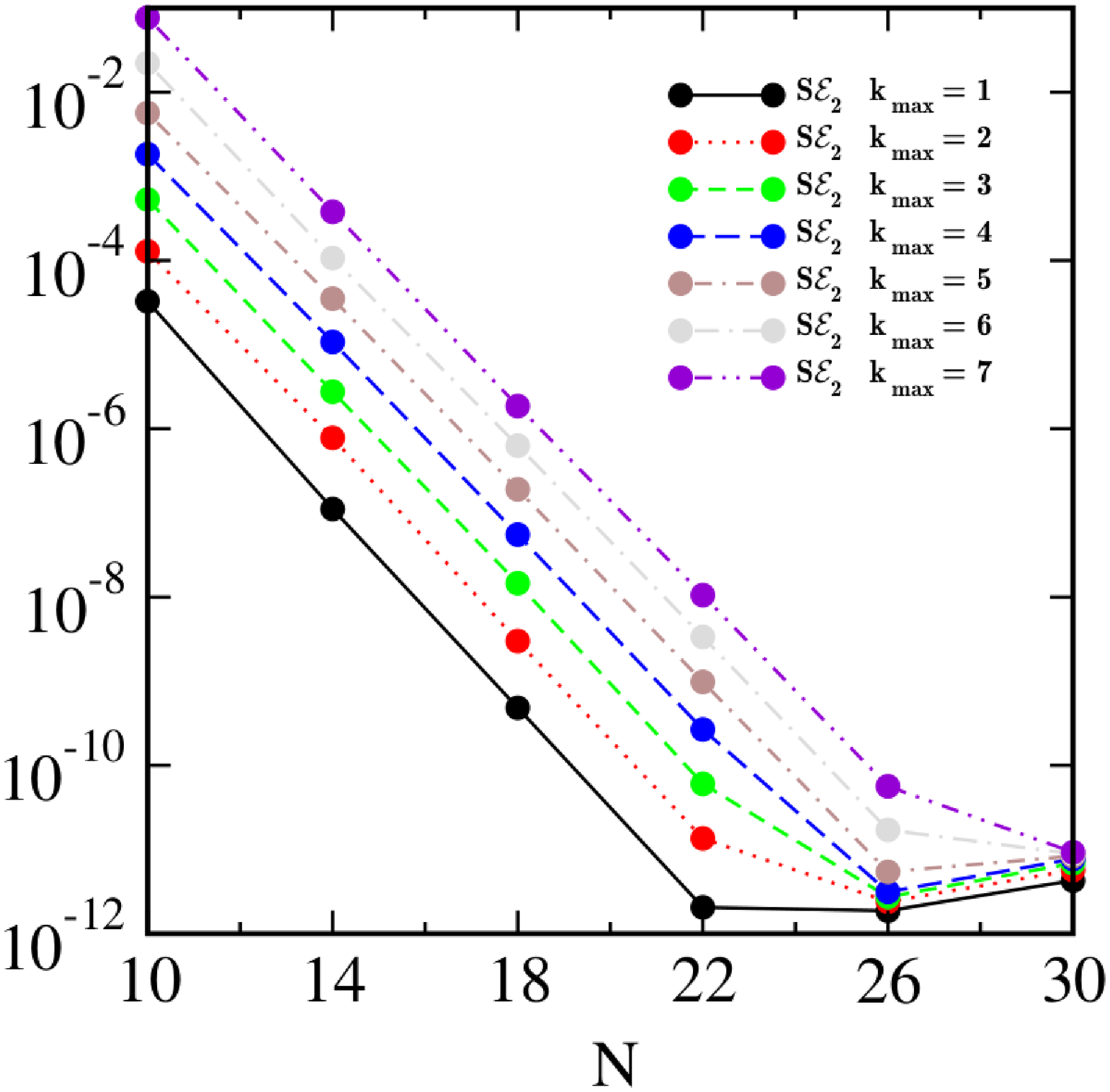}\hfill
  \includegraphics[width=0.31\textwidth]{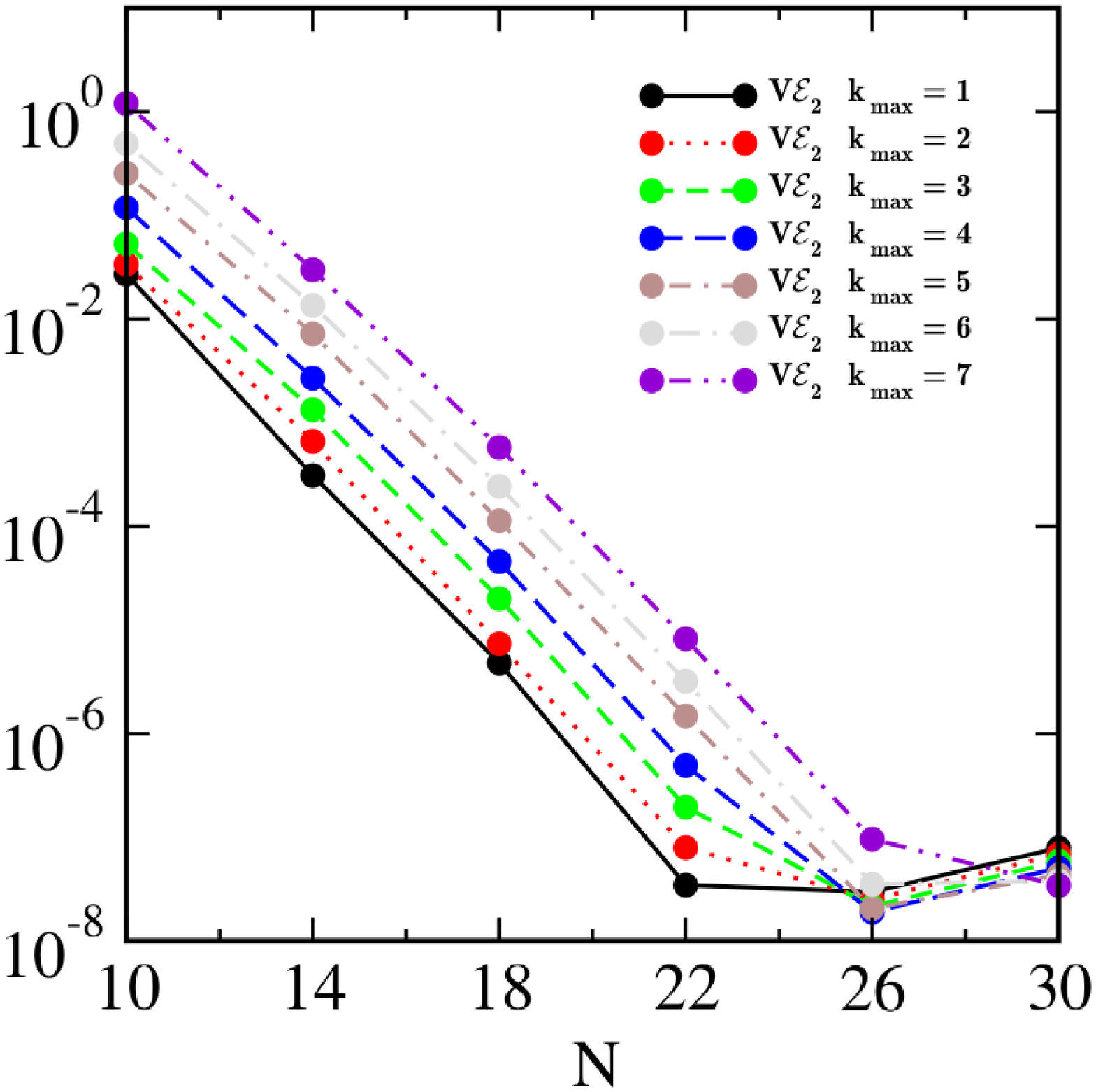}\hfill
  \includegraphics[width=0.31\textwidth]{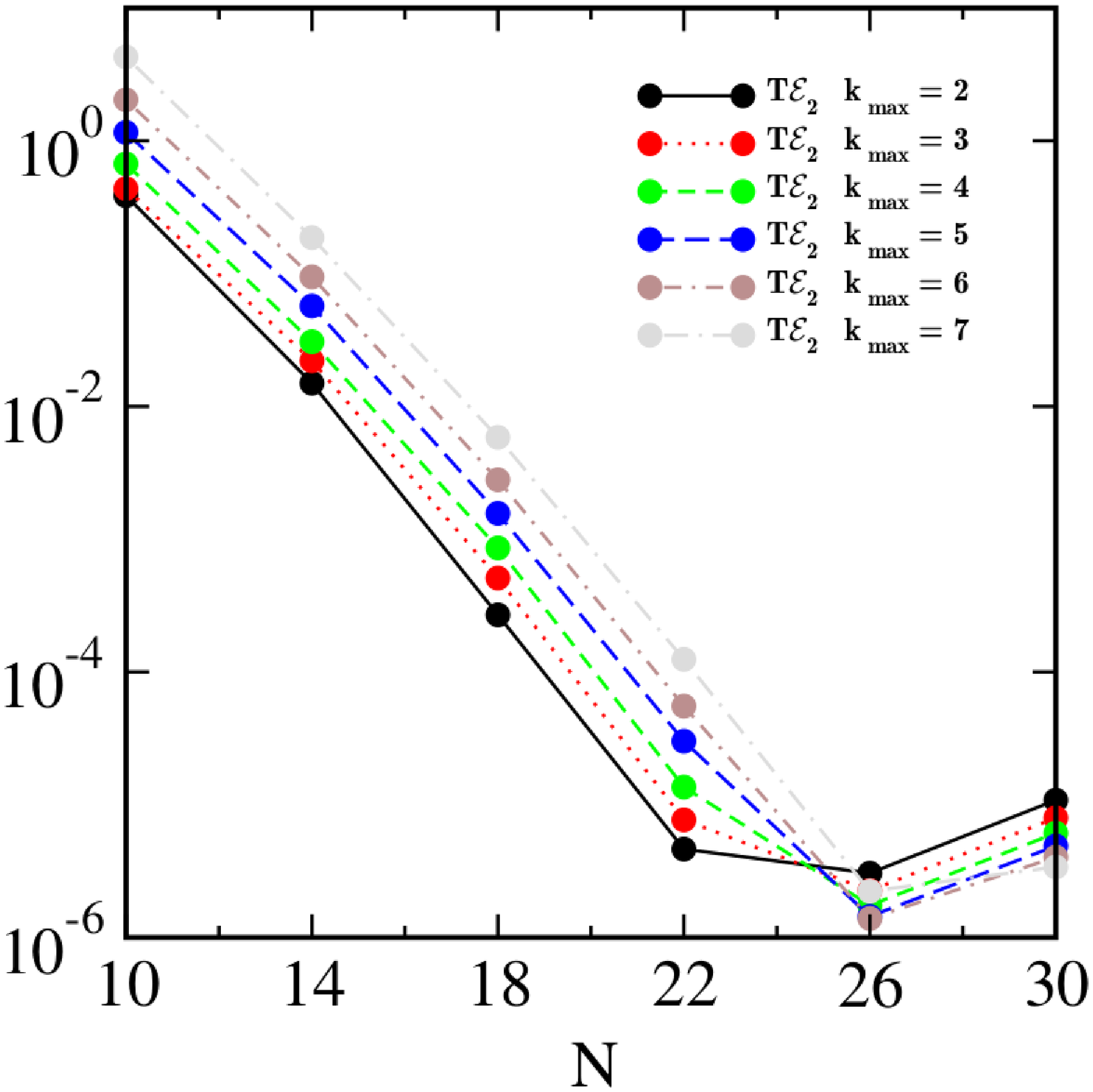}

  \caption{\label{f:EigenRes} Values of $S\mathcal{E}_2$, $V\mathcal{E}_2$, 
    and $T\mathcal{E}_2$, the composite scalar, vector, and tensor Laplace 
    operator eigenfunction residuals defined in
    Eqs.~(\ref{e:SE2})--(\ref{e:TE2}), respectively. 
    These values are plotted as functions of the number of grid points $N$
    used in each dimension of each of the eight computational
    sub-domains.}
\end{figure}

Figure~\ref{f:EigenRes} illustrates the
composite Laplace operator eigenvalue residuals defined in
Eqs.~(\ref{e:SE2})--(\ref{e:TE2}) for our numerical scalar, vector and
tensor harmonics.  Each curve represents one of the
composite residual norms for a fixed value of $k_\mathrm{max}$ as a
function of the numerical resolution $N$. These plots show that our
numerical methods converge exponentially in the numerical resolution
$N$ (which is typical for pseudo-spectral methods), and they also
illustrate how the residuals depend on the order of the harmonics,
$k_\mathrm{max}$.  The process of evaluating the derivatives of fields
numerically is always a significant source of error in any
calculation. Evaluating the covariant Laplace operator eigenvalue
residuals requires two numerical derivatives of the $\mathbb{S}^3$ harmonics.
The values of these residuals are therefore expected to be larger (for
given $k_\mathrm{max}$ and $N$) than those requiring only one, or no
numerical derivatives at all.

We also define composite residuals to measure how well the
divergence and trace identities are satisfied:
\begin{eqnarray}
  (V\mathcal{D}_2)^2 &=&
\sum_{A=0}^2\sum_{k=k^{V(A)}_\mathrm{min}}^{k_\mathrm{max}}
  \sum_{\ell=\ell^{V(A)}_\mathrm{min}}^k\sum_{m=0}^\ell  
\frac{\bigl(\| \mathcal{D}^{k\ell m}_{(A)}\|_2\bigr)^2}
{3N_{k\ell m}^\geq-2k_\mathrm{max}-3},
  \label{e:VD2} \\[1em]
  (T\mathcal{D}_2)^2 &=&
\sum_{A=0}^5\sum_{k=k^{T(A)}_\mathrm{min}}^{k_\mathrm{max}}
  \sum_{\ell=\ell^{T(A)}_\mathrm{min}}^k\sum_{m=0}^\ell 
  \frac{\bigl(\|\mathcal{D}^{k\ell m}_{(A)\,a}\|_2\bigr)^2}
       {6N_{k\ell m}^\geq-8k_\mathrm{max}-12},
  \label{e:TD2}\\[1em]
  (\mathcal{T}_2)^2 &=&
  \sum_{A=0}^5\sum_{k=k^{T(A)}_\mathrm{min}}^{k_\mathrm{max}}
  \sum_{\ell=\ell^{T(A)}_\mathrm{min}}^k\sum_{m=0}^\ell 
\frac{\bigl(\|\mathcal{T}^{k\ell m}_{(A)}\|_2\bigr)^2}
     {6N_{k\ell m}^\geq-8k_\mathrm{max}-12}.
  \label{e:Tr2} \\ \nonumber
\end{eqnarray}

\begin{figure}
  \centering
  \includegraphics[width=0.31\textwidth]{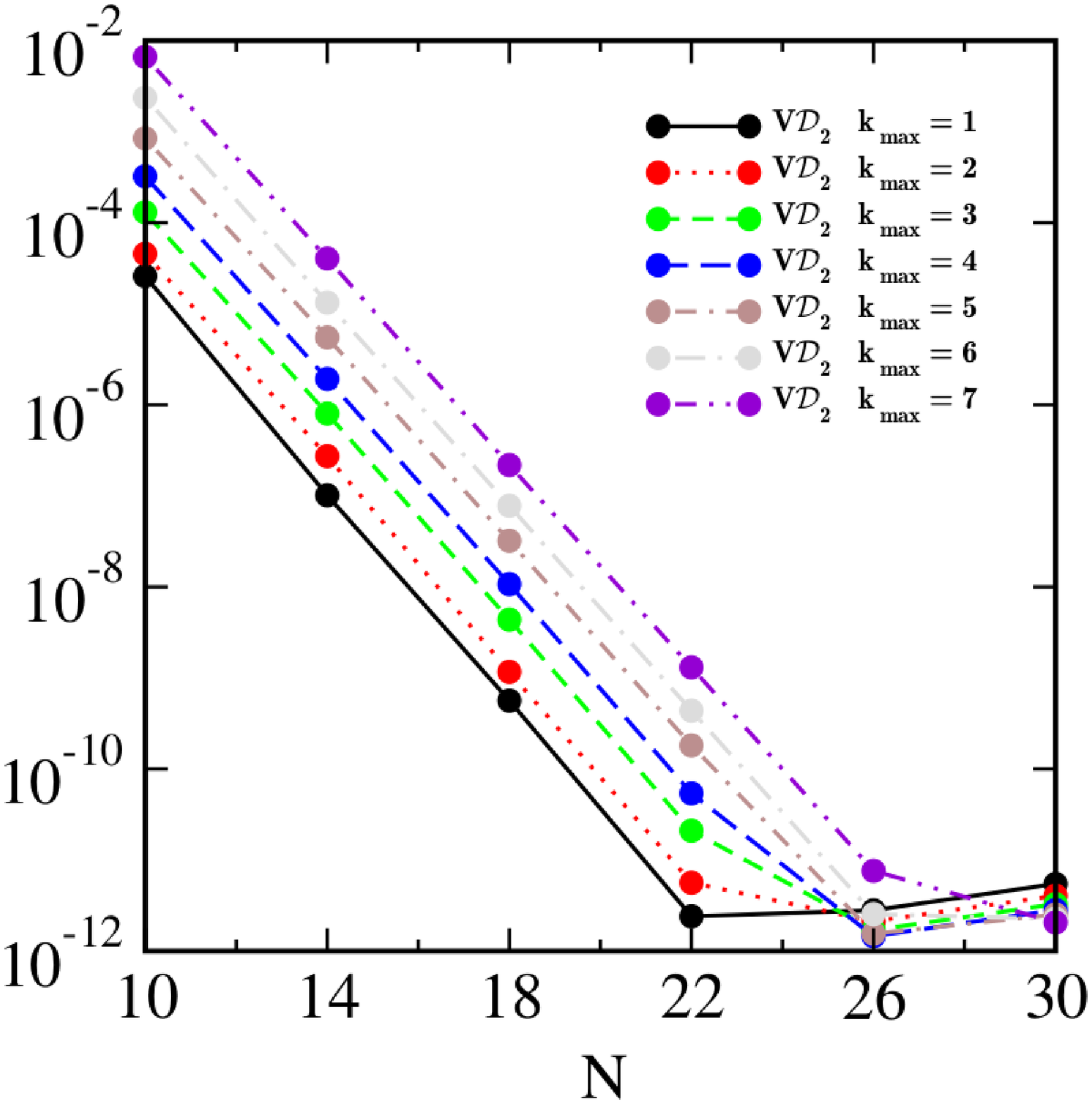}\hfill
  \includegraphics[width=0.31\textwidth]{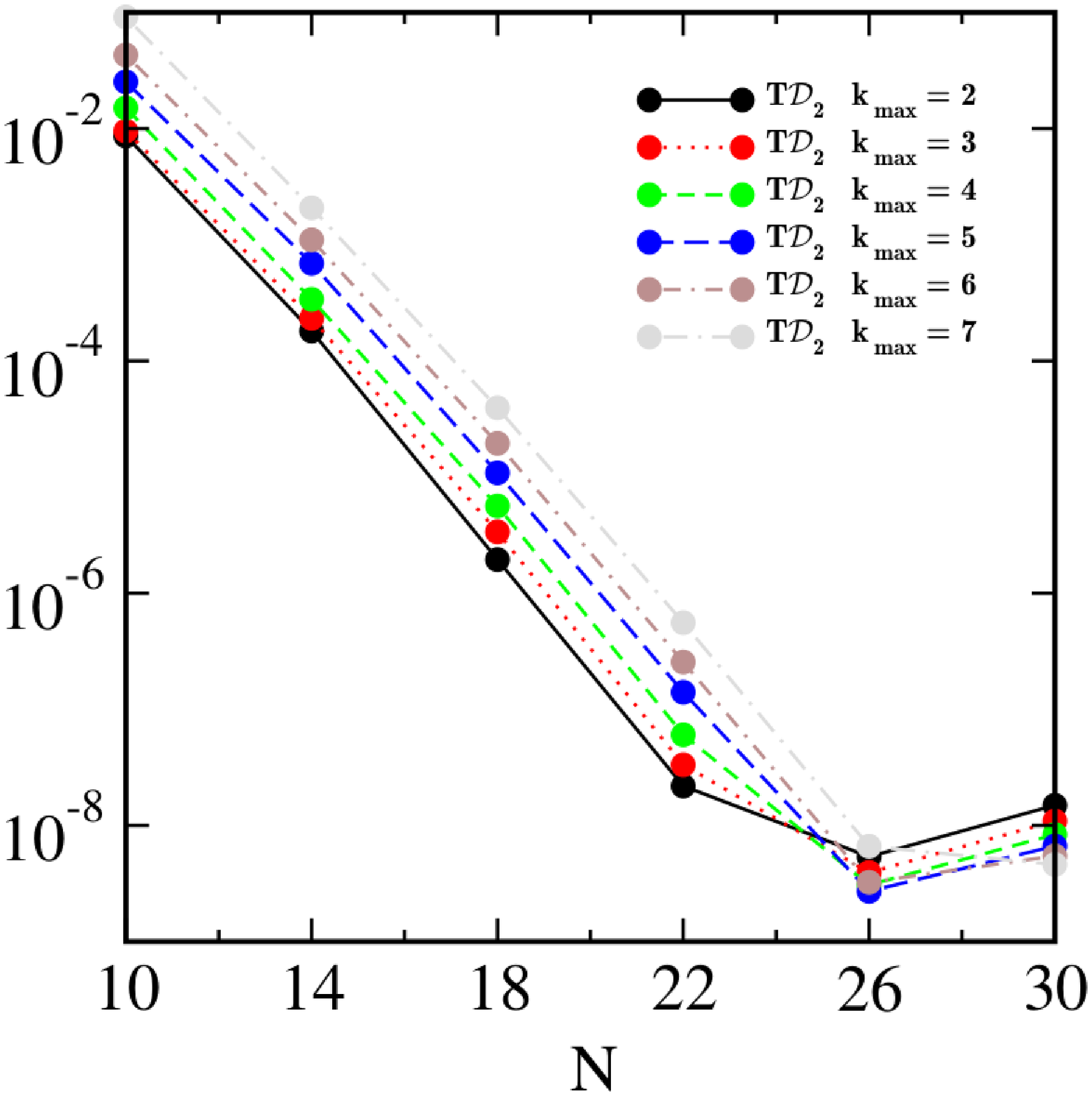}\hfill
  \includegraphics[width=0.31\textwidth]{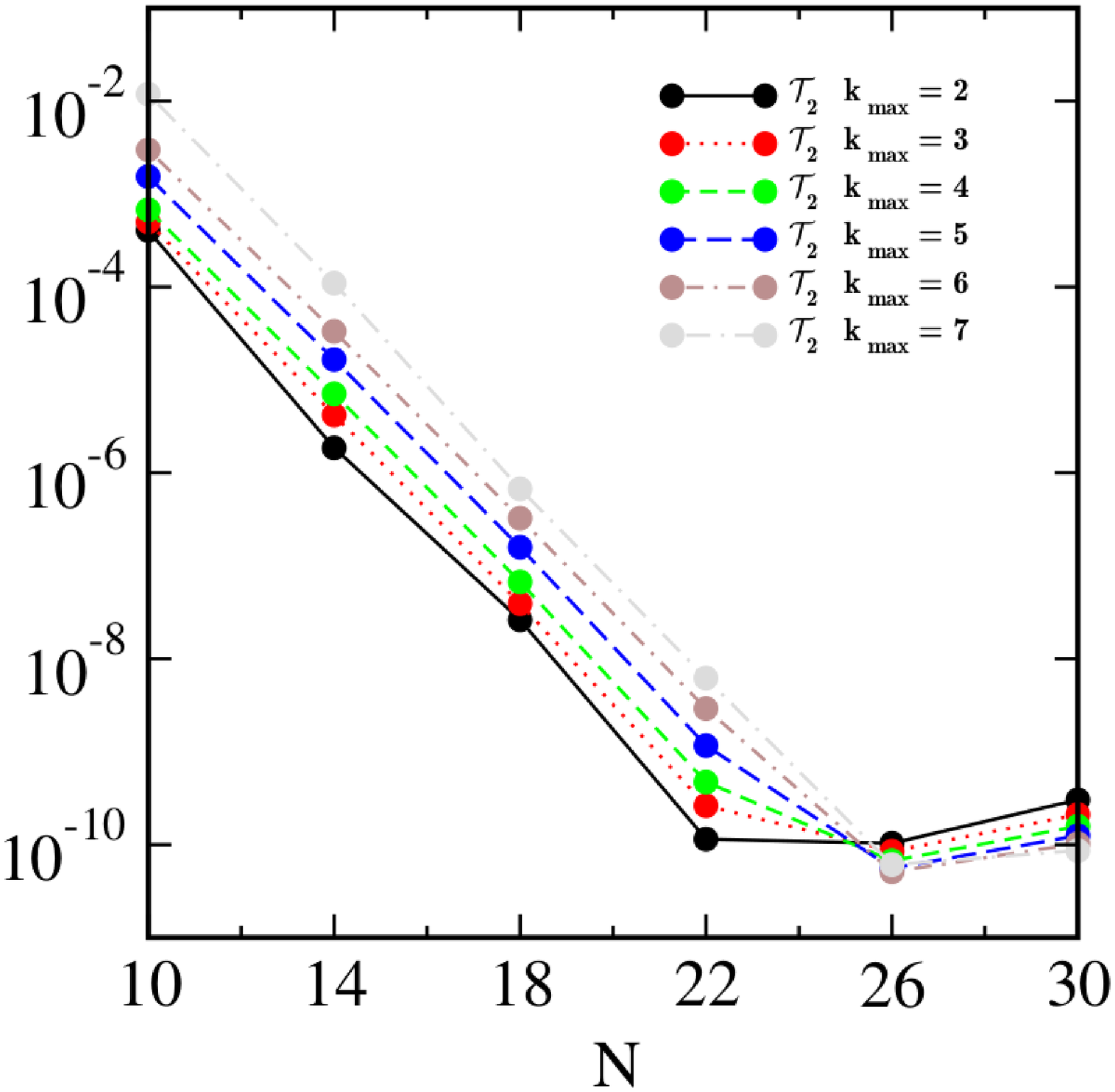}

  \caption{\label{f:DivTraceRes} Values of $V\mathcal{D}_2$, 
    $T\mathcal{D}_2$, and $\mathcal{T}_2$, the
    composite vector and tensor divergence residuals, and tensor trace 
    residual, defined in Eqs.~(\ref{e:VD2})--(\ref{e:Tr2}), respectively.
    These values are plotted as functions of the number of grid points $N$
    used in each dimension of each of the eight computational
    sub-domains.}
\end{figure}

Figure~\ref{f:DivTraceRes} illustrates the
composite divergence and trace residuals defined in
Eqs.~(\ref{e:VD2})--(\ref{e:Tr2}). Each curve in each figure
represents one of the composite residual norms for a fixed value of
$k_\mathrm{max}$ as a function of the numerical resolution $N$.  These
plots illustrate how the values of these residuals depend both on
$k_\mathrm{max}$ and on the numerical resolution $N$ used to evaluate
them.

Finally we define composite residuals that measure how well the
ortho-normality residuals are satisfied:
\begin{eqnarray}
  &&(S\mathcal{O}_2)^2 =
\sum_{k=0}^{k_\mathrm{max}}\sum_{\ell=0}^k\sum_{
    m=-\ell}^\ell\sum_{k'=0}^{k_\mathrm{max}}\sum_{\ell'=0}^{k'}\sum_{
    m'=-\ell'}^{\ell'}
\frac{\left|S\mathcal{O}^{k\ell m}_{k'\ell'm'}\right|^2}{N_{k\ell m}^2}
,\label{e:SO2}\\
&&(V\mathcal{O}_2)^2 = \sum_{A=0}^2\sum_{B=0}^2
\sum_{Vk\ell m}^{k_\mathrm{max}}\sum_{Vk'\ell' m'}^{k_\mathrm{max}}
 \frac{\left|V\mathcal{O}^{(A)k\ell m}_{(B)k'\ell'm'}\right|^2}
      {\left(3N_{k\ell m}-2k_\mathrm{max}-3\right)^2},\qquad
      \label{e:VO2}
\end{eqnarray}
\begin{eqnarray}
  &&(T\mathcal{O}_2)^2 =\sum_{A=0}^5\sum_{B=0}^5
  \sum_{Tk\ell m}^{k_\mathrm{max}}\sum_{Tk'\ell' m'}^{k_\mathrm{max}}
 \frac{\left|T\mathcal{O}^{(A)k\ell m}_{(B)k'\ell'm'}\right|^2}
     {\left(6N_{k\ell m}-10k_\mathrm{max}-15\right)^2},\qquad
     \label{e:TO2}
\end{eqnarray}
where $\sum_{Vk\ell m}^{k_\mathrm{max}}$ and $\sum_{Tk\ell m}^{k_\mathrm{max}}$ are  defined as
\begin{eqnarray}
\sum_{Vk\ell m}^{k_\mathrm{max}} &=&  \sum_{k=k^{V(A)}_\mathrm{min}}^{k_\mathrm{max}}
 \sum_{\ell=\ell^{V(A)}_\mathrm{min}}^k\sum_{m=-\ell}^\ell,\\
  \sum_{Tk\ell m}^{k_\mathrm{max}} &=&  \sum_{k=k^{T(A)}_\mathrm{min}}^{k_\mathrm{max}}
 \sum_{\ell=\ell^{T(A)}_\mathrm{min}}^k\sum_{m=-\ell}^\ell,
\end{eqnarray}
and where $N_{k\ell
  m}=(k_\mathrm{max}+1)(k_\mathrm{max}+2)(2k_\mathrm{max}+3)/6$ is the
number of $k\ell m$ triplets with $-\ell\leq m \leq \ell$.  The terms
$2k_\mathrm{max}+3$ and $10k_\mathrm{max}+15$ that appear in
Eqs.~(\ref{e:VO2}) and (\ref{e:TO2}) respectively represent the number
of terms excluded in these sums by the lower bounds $k\geq
k^{T(A)}_\mathrm{min}$ and $\ell\geq \ell^{T(A)}_\mathrm{min}$.

\begin{figure}
  \centering
  \includegraphics[width=0.31\textwidth]{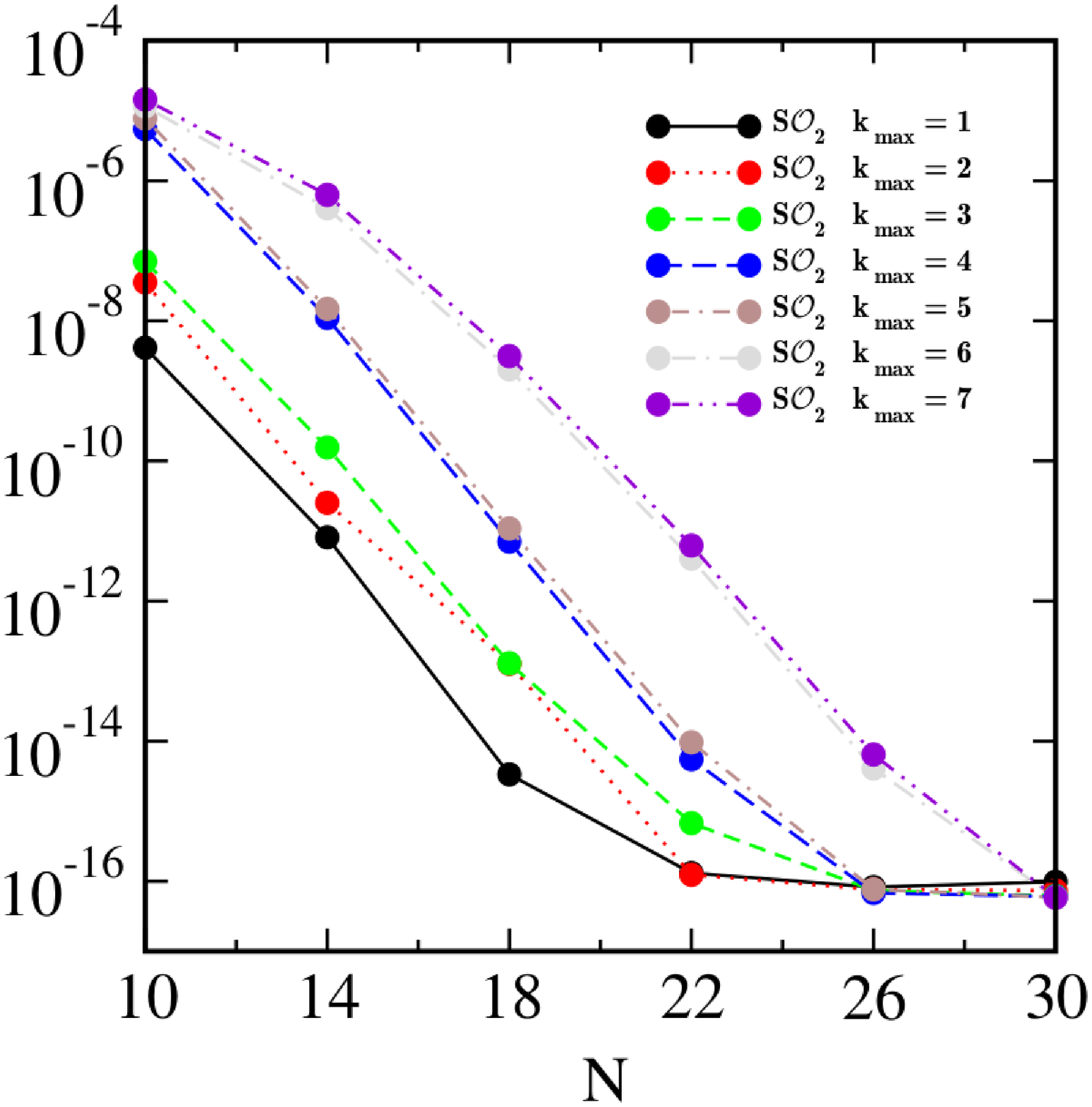}\hfill
  \includegraphics[width=0.31\textwidth]{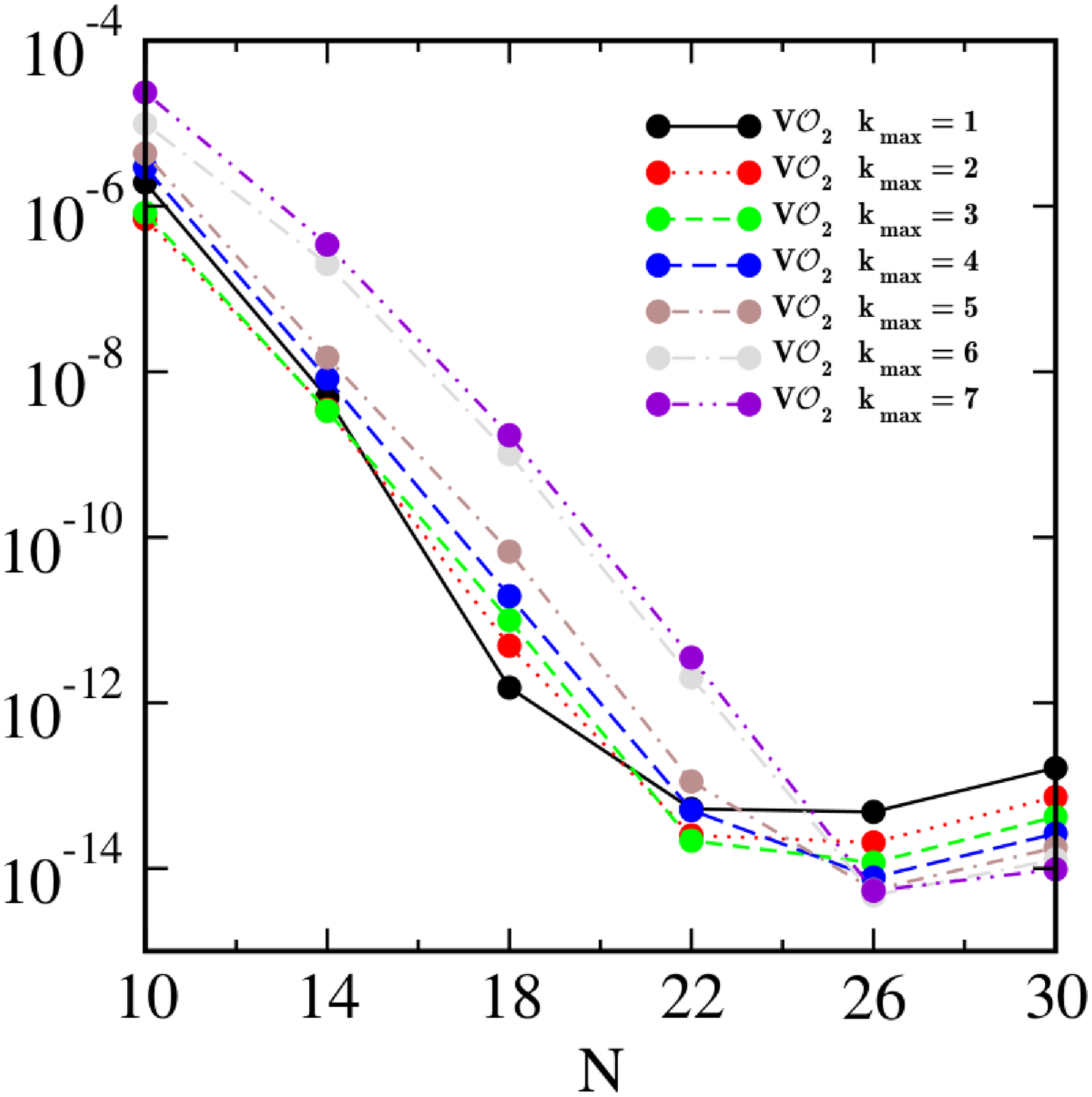}\hfill
  \includegraphics[width=0.31\textwidth]{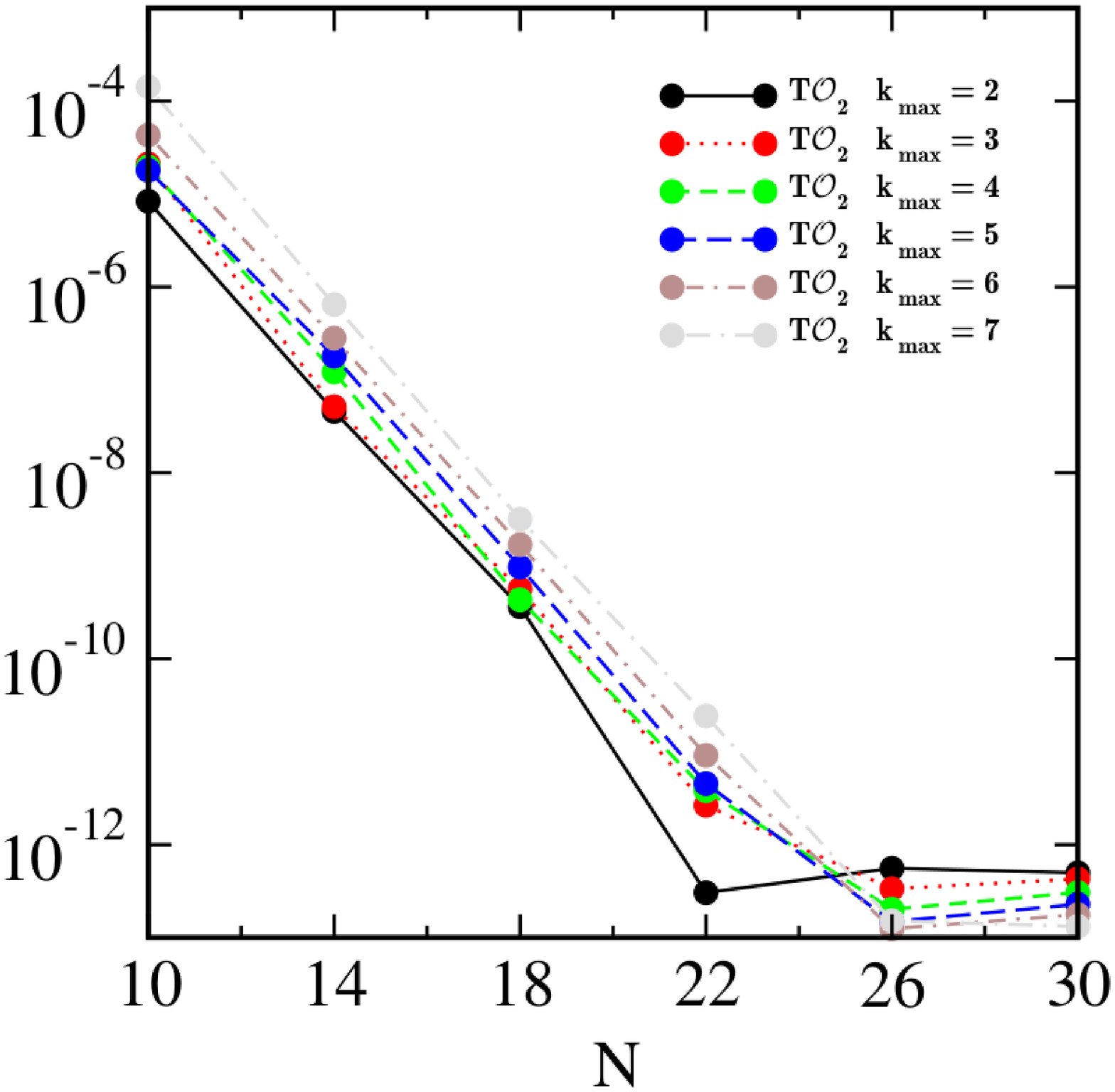}

  \caption{\label{f:OrthoRes} Values of $S\mathcal{O}_2$, $V\mathcal{O}_2$,
    and $T\mathcal{O}_2$, the
    composite scalar, vector, and tensor ortho-normality residuals defined in
    Eqs.~(\ref{e:SO2})--(\ref{e:TO2}), respectively.
    These values are plotted as functions of the number of grid points $N$
    used in each dimension of each of the eight computational
    sub-domains.}
\end{figure}

Figure~\ref{f:OrthoRes} illustrates the
composite ortho-normality residuals defined in
Eqs.~(\ref{e:SO2})--(\ref{e:TO2}) for our numerical scalar, vector and
tensor harmonics. Each curve represents one of the
composite ortho-normality residual norms for a fixed value of
$k_\mathrm{max}$ as a function of the numerical resolution $N$.  These
plots illustrate how the values of these residuals depend both on
$k_\mathrm{max}$ and on the numerical resolution $N$ used to evaluate
them.  These composite ortho-normality residuals depend on numerical
integrals of the $\mathbb{S}^3$ harmonics, but not on their numerical
derivatives.  Consequently these residuals are expected to be much
smaller (for fixed $k_\mathrm{max}$ and $N$) than the Laplace operator
eigenfunction residuals and the divergence residuals.


\section{Summary}
\label{s:Summary}
We have summarized the useful properties of the
scalar, vector, and tensor harmonics on the three-sphere, and we
have presented a new notation that unifies,
simplifies and clarifies the analytical
expressions for these harmonics. As such, these expressions are in a
form that is well-suited for straightforward numerical
implementation. We have performed numerical tests
of the harmonics computed in this way, and have presented
results that demonstrate the accuracy and
convergence of the methods.

\begin{acknowledgements}
We thank the Center for Computational Mathematics at the University of
California at San Diego for providing access to their computer cluster
on which all the numerical tests reported in this paper were
performed.  LL's research was supported in part by grants PHY 1604244
and DMS 1620366 from the National Science Foundation to the University
of California at San Diego.  FZ's research was partially supported by
the NSFC grants 11503003 and 11633001, Strategic Priority Research
Program of the Chinese Academy of Sciences Grant No. XDB23000000, the
Fundamental Research Funds for the Central Universities Grant
2015KJJCB06, and a Returned Overseas Chinese Scholars Foundation
grant.
\end{acknowledgements}


\bibliographystyle{spphys}
\bibliography{../References/References}

\end{document}